\documentclass[manuscript]{aastex}

\slugcomment{Not to appear in Nonlearned J., 45.}

\shorttitle{The Outer Disk of the LMC}
\shortauthors{Carrera et al.}

\begin{document}


\title{Metallicities, Age--Metallicity relationships, and Kinematics of
Red Giant Branch Stars in the Outer Disk of the Large Magellanic Cloud\footnote{Based on observations collected at the European Southern Observatory,
Chile, within the observing programs 074.B-0474 and 082.B-0900}}


\author{R. Carrera, C. Gallart and A. Aparicio}
\affil{Instituto de Astrof\'{\i}sica de Canarias, Spain}
\affil{Departamento de Astrof\'{\i}sica, Universidad de La Laguna, Spain}
\email{rcarrera@iac.es}

\and

\author{E. Hardy}
\affil{National Radio Astronomy Observatory, Chile\footnote{The National Radio Astronomy Observatory is a facility of the National
Science Fundation operated under cooperative agreement by Associated
Universties, Inc.}}
\affil{Departamento de Astronom\'{\i}a, Universidad de Chile, Chile\footnote{Adjoint Professor}}


\begin{abstract}
The outer disk of the Large Magellanic Cloud (LMC) is studied in order to unveil clues about its formation and evolution. Complementing our previous studies in innermost fields (3 kpc$\lesssim$R$\lesssim$7 kpc), we obtained deep color magnitude diagrams in 6 fields with galactocentric distances  from 5.2 kpc to 9.2 kpc and different azimuths. The comparison with isochrones shows that while the oldest population is approximately coeval in all fields, the age of the youngest populations increases with increasing radius. This agrees with the results obtained in the innermost fields. Low-resolution spectroscopy in the infrared \ion{Ca}{2} triplet region has been obtained for about 150 stars near the tip red giant branch in the same fields. Radial velocities and stellar metallicities have been obtained from these spectra. The metallicity distribution of each field has been analyzed together with those previously studied. The metal content of the most metal-poor objects, which are also the oldest according to the derived age-metallicity relationships, is similar in all fields independently of the galactocentric distance. However, while the metallicity of the most metal-rich objects measured, which are the youngest ones, remains constant in the inner 6 kpc, it decreases with increasing radius from there off. The same is true for the mean metallicity. According to the derived age-metallicity relationships, which are consistent with being the same in all fields, this result may be interpreted as an outside-in formation scheme in opposition with the inside-out scenario predicted by $\Lambda$CDM cosmology for a galaxy like the LMC. The analysis of the radial velocities of our sample of giants shows that they follow a rotational cold disk kinematics. The velocity dispersion increases as metallicity decreases indicating that the most metal-poor/oldest objects are distributed in a thicker disk than the most metal-rich/youngest ones in agreement with the findings in other disks such as that of the Milky Way. They do not seem to be part of a hot halo, if one exists in the LMC.
\end{abstract}

\section{Introduction}

The formation and evolution of stellar disks in galaxies are still not well understood processes in spite of the observational and theoretical effort developed in the last years \citep[see][for a review]{vanderkruit2011}. Simulations based on $\Lambda$CDM predict that the inner parts of disks form first and then they grow up to reach their present-day size \citep[e.g.][]{white1991,mo1998,roskar2008}. According to this scenario, so-called inside-out, it is expected that the youngest populations are located in the outermost regions. Therefore, to understand the formation and evolution of disks, it is important to have information about the spatial distribution of their different stellar populations in a large range of galactocentric distances. In particular, their outskirts are key since they might retain a fossil record of the disk formation due to their long dynamical timescales.

From an observational point of view, the study of the outer disks is difficult due to their low-surface brightness at the level of, or even bellow, the background. The behaviour of the star formation rate with radius can be studied with surface photometry in galaxies within a few Mpc \citep[e.g.][]{dejong1996,taylor2005}. In general, these investigations found that massive early-type spirals are bluer in their outer disk \citep[e.g.][]{trujillo2005}, in agreement with the inside-out scenario although these disks are mainly populated by old ($\sim$10 Gyr) old stars \citep[e.g.][]{sanchezblazquez2011}. Disks in late-type spirals and irregular galaxies have more puzzling behaviors \citep[e.g.][]{dejong1996,gadotti2001,taylor2005,hunter2006}. Some of them follow the same trend observed in early-type spirals. However, others become redder outward which is interpreted as the youngest populations being located in the innermost regions, since no galaxy formation model predicts positive metallicity gradients. This suggests an outside-in scenario in the way predicted by \citet{stinson2009} for less massive galaxies and observed by \citet{hidalgo2011} for less massive (M$\lesssim$10$^9$ M$_\odot$) galaxies.

Nearby galaxies, in which their stars can be resolved, offer the opportunity to study galactic disks in more detail. The observations in the two most massive spirals in the Local Group, M31 and the Milky Way, agree quite well with the predictions of the inside-out scenario. M31 does not show a clear metallicity gradient \citep{ibata2005} while in the Milky Way a smooth gradient is observed which was more sharp in the past \citep[e.g.][]{carrerapancino2011}. There is another spiral in the Local Group, M33, of later type and much less massive than the previous ones. This galaxy shows a truncation in its radial brightness surface profile \citep{ferguson2007} which is also observed in other galaxies \citep[e.g.][]{trujillo2005}. Although still not well understood, this truncation can be originated during the collapse of the protogalactic cloud \citep{vanderkruit1987} or can be related to different star formation thresholds at different galactocentric distances \citep[e.g.][]{schaye2004}. The scale-length of M33 increases with time within the break radius \citep{williams2009} and steepens and even reverse from there out \citep{barker2007}. These authors interpreted this result in the context of the inside-out scenario: the inner disk would have been grown from the inner regions, while secular processes radially redistribute stars causing an inflection point in the stellar age gradient at the break radius.

There is a fourth disk galaxy in the Local Group, the Large Magellanic Cloud (LMC). Sometimes considered an irregular, it is more properly classified as a Barred Magellanic Spiral (SBm). This kind of objects are frequent through the Universe and are characterised by a mass between that of large and dwarf galaxies, and the presence of a bar and a disk containing a single spiral arm. The proximity of the LMC allows us to investigate it using both spectroscopy and photometry of individual stars. Therefore, we can obtain information about the age and/or chemical abundances spatial distribution and about kinematics. Moreover, the LMC contains the only resolved bar in the Local Group and it forms an interacting system together with the Small Magallanic Cloud and the Milky Way, although their history of past interactions are still not well understood \citep[e.g.][]{besla2007,besla2010}.

The LMC inner disk\footnote{In this paper we call ``inner disk'' the region within $R\sim4.5$ kpc, and ``outer disk'' the region from that distance off. Note
that at the distance of the LMC, 1 kpc corresponds to 1$\fdg$12 on
the sky \citep{vandermarel2001}} ($R<4.5$ kpc) has been the subject of most work about its stellar population \citep[e.g.][]{bertelli1992,olsen1999,holtzman1999,cole2000,castro2001,smeckerhane2002,cole2005,pompeia2008,harris2009}.
However, the LMC disk extends at least to 12-14 kpc \citep{majewski2009,saha2010}. The lack of works in the outer LMC do not allow to study radial trends in detail. Studies covering large galactocentric distances use stellar clusters as test particles \citep[e.g.][]{olszewski1991,grocholski2006,sharma2010}. They did not find neither age nor metallicity gradients. The spatial distribution of asymptotic giant branch stars (AGB) in the field of the LMC has been also studied using the ratio between the number of carbon and oxygen-rich stars, traditionally denoted as C/M. Two works have used this indicator reaching contradicting conclusions for the inner disk. \citet{cioni2009} found that the C/M ratio increases rapidly with increasing distance which is interpreted as a metallicity gradient with the outskirts being more metal-poor. However, using the same data \citet{feast2010} found that the C/M ratio is nearly constant in the inner 4 kpc,  from there off the ratio decreases. Since carbon-rich stars are rare in old populations this result is interpreted as the outer disk stars are older. Unfortunately, high-resolution spectroscopy
studies at different galactocentric distances, which would allow investigation
of the possible presence of abundance gradients of some key chemical
species, are not available.

Our group is involved in a major effort to study the LMC using both photometric and spectroscopic observations. \citet[][hereafter Paper I]{gallart2008} studied four fields located at different galactocentric distances (3 kpc $<$R$<$ 7 kpc)\footnote{The galactocentric distances used in Paper I are different from those
used here and in Paper II, because they have been calculated assuming
different centres. While the dynamical center of the Carbon stars
was used here, which better represents the center of the older populations, paper I used the center of the Bar. The distance between
both is about 1 kpc.}. They found that the age of the youngest populations in each field increases with galactocentric distance. In fact, a more detailed analysis of these fields by \citet{meschin2010} found a gradual decrease of the
ratio of young to old populations with radius. \citet[hereafter Paper II]{carrera2008a}
derived ages and metallicities for about 100 red giant branch (RGB)
stars confirmed as LMC members from their radial velocities in each
of these fields using the infrared \ion{Ca}{2}
triplet (CaT) lines. The three inner fields have a similar mean metallicity, but that of the outermost
one is a factor of two more metal-poor (see Paper II for details). This
is explained by the lack of the youngest stars in the latter field
which are also more metal-rich. These results point toward an outside-in formation of the LMC disk, on contrary to the inside-out scenario usually predicted  for a galaxy with a mass similar to that of the LMC. The observed behaviour is more like that expected for less massive galaxies \citep{stinson2009,hidalgo2011}. 

In this paper, we extend our study of the LMC disk at larger galactocentric distances (R$>$7 kpc). For this purpose we have obtained deep CMDs and low resolution spectroscopy
for 6 new fields spread over the LMC disk. Their spatial distribution
and color-magnitude diagrams (CMD) are described in Sections \ref{spatial_distribution} and \ref{cmd}, respectively. The
spectroscopic observations and data reduction, together with the tools
used to derive radial velocities, metallicities, and ages are described
in Section \ref{datareduction}. The metallicity distribution and
age-metallicity relation (AMR) of each field are presented in Sections \ref{metaldistributions},
and \ref{amr}, respectively. The kinematics of our RGB sample is
analysed in Section \ref{kinematics}. Finally, the implications of
the main results of this work are discussed in Section \ref{sec_discuss}
and summarised in Section \ref{sec_summa}.

\section{Spatial Distribution of Observed Fields}\label{spatial_distribution}

\begin{figure}
\epsscale{1} 
\includegraphics[angle=0,scale=.50]{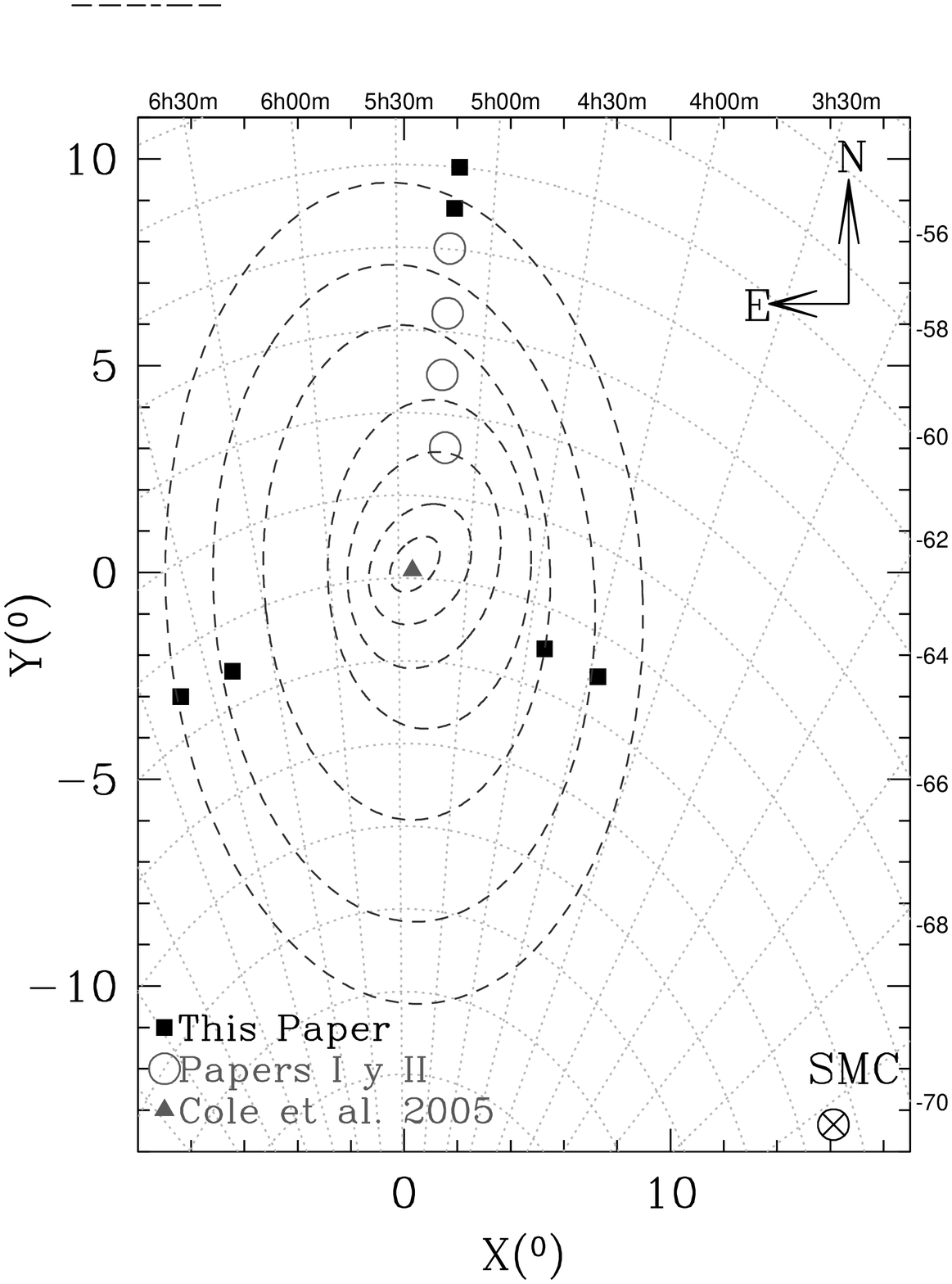}
\caption{Spatial distribution of the fields studied in this paper (filled squares)
together with those studied in Papers I and II (open circles).
The field studied by \citet{cole2005} in the bar is also marked (filled
triangle). As reference, the isopleths of the RGB and
AGB stars derived by \citet{vandermarel2001}
have been superimposed with semimajor-axis of 1\arcdeg, 2\arcdeg,
3\arcdeg, 4\arcdeg, 6\arcdeg, 8\arcdeg, and 10\arcdeg. The position
of the SMC is also shown in the bottom-right corner. Right ascension
and declination are labeled on top and right axis, respectively.\label{fig1}}
\end{figure}

We have selected 6 fields in addition to the ones presented in Papers I and
II. They are shown in Figure \ref{fig1} together with the near infrared
isopleths from \citet{vandermarel2001} for reference. Two of the
new fields presented in this paper are situated at either side of
the LMC bar, in the East-West direction, toward the SMC (West)
and the Galactic center (East). With them we aim to investigate the
possible influence of the SMC in the LMC's disk evolution. The remaining two fields
are located northward of the bar, as those studied in Paper II, but
further out than them. These fields were selected to study the outermost
regions of the LMC disk and are key to investigate the presence of
a metallicity gradient.

Equatorial -right ascension and declination- coordinates over the
celestial sphere for observed fields, named according to their right
ascension and declination as in Paper I, are listed in Table \ref{tbl1}.
Angular -distance and position angle- coordinates have been computed
using the equations described by \citet{vandermarel2001} considering
the dynamical center of the carbon stars as origin \citep[$\alpha_{2000}=5^h27\fm6$ and $\delta_{2000}=-69\fdg87$; ][]{vandermarel2002}
are also listed in Table~\ref{tbl1}. Distances computed
in this way might not be actual physical distances because they do not take into
account the viewing angles of the disk: inclination, $i$, and the
major--axis position angle, $P.A._{maj}$, which are needed to deproject
the observed morphology. We deproject the angular distances
of our fields $R_{proj}$ following \citet{cioni2009}, assuming $i=34\fdg4\pm6\fdg2$
and $P.A._{maj}=189\fdg3\pm1\fdg4$ \citep{vandermarelcioni2001}.
These values, listed in the last column of Table \ref{tbl1}, will
be used throughout this work.

\section{Color-Magnitude Diagrams}\label{cmd}

$V$- and $I$-band images for the northern fields, and $B$- and $R$-band images for the others, were obtained with WFI at the 2.2 m telescope
at La Silla Observatory, Chile. The WFI camera is a mosaic of 4 $\times$
2 of 2048 $\times$ 4096 CCDs covering a total area of 34\arcmin~$\times$
33\arcmin~with a scale of 0.238 pixel$^{-1}$. The images were bias and flat-field corrected in the usual way with the IRAF\footnote{IRAF is distributed by the National Optical Astronomy Observatories,which is operated by the Association of Universities for Research
in Astronomy, Inc., under cooperative agreement with the National
Science Foundation.} \textit{MSCRED} package. Profile-fitting photometry was obtained with the DAOPHOT/ALLFRAME suite of codes \citep{stetson1994} and calibrated to the standard system using observations of several \cite{stetson2000} fields obtained in the same run.

\begin{figure}
\includegraphics[angle=0,scale=0.75]{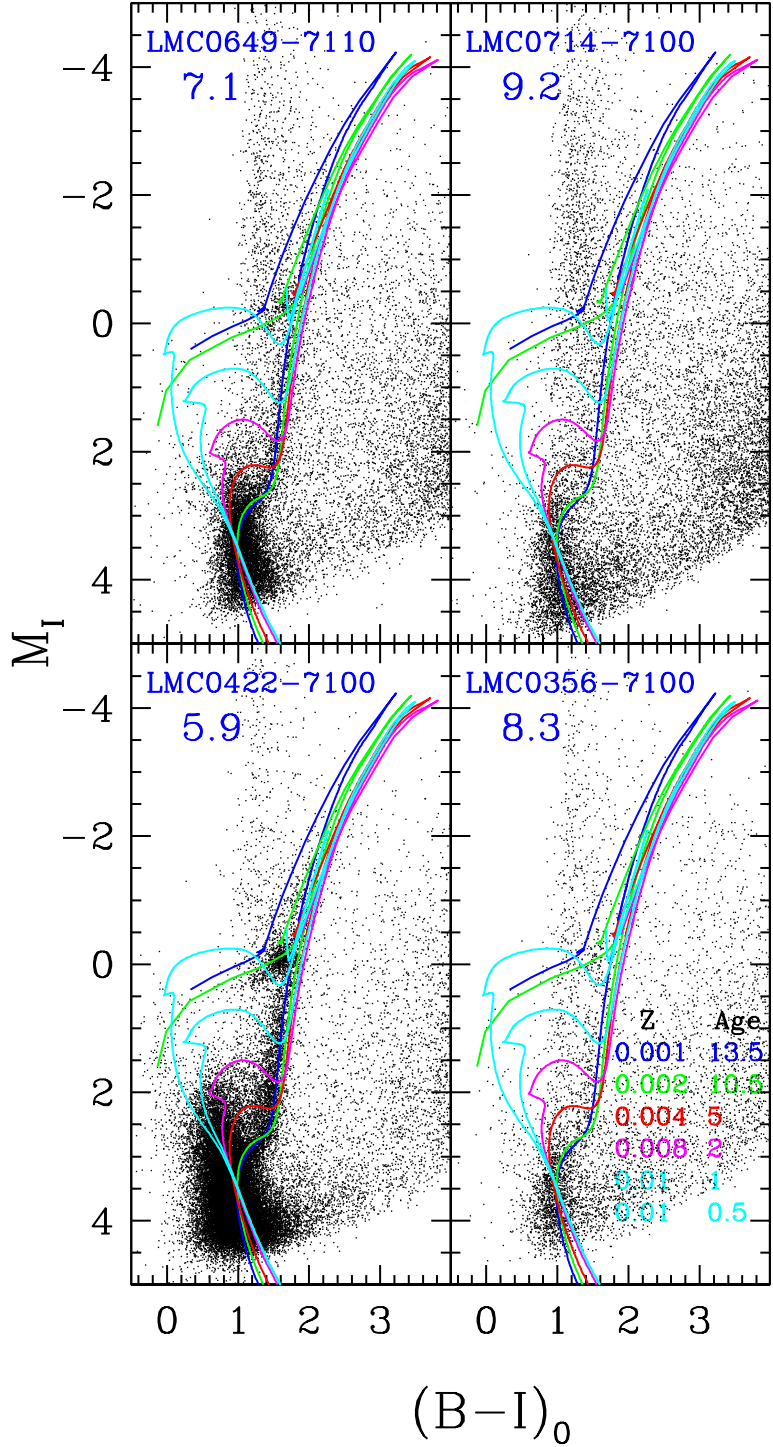}
\caption{[$(B-I)_{0}$,$M_{I}$] CMDs for the eastern (top) and western
(bottom) fields. Isochrones with ages and metallicities as labeled,
and a zero-age horizontal branch by \citet{pietrinferni2004} have
been superimposed. The identification of each field together with its galactocentric distance, in kpc, are labeled in top-left corner.\label{fig2}}
\end{figure}

\begin{figure}
\includegraphics[angle=0,scale=.75]{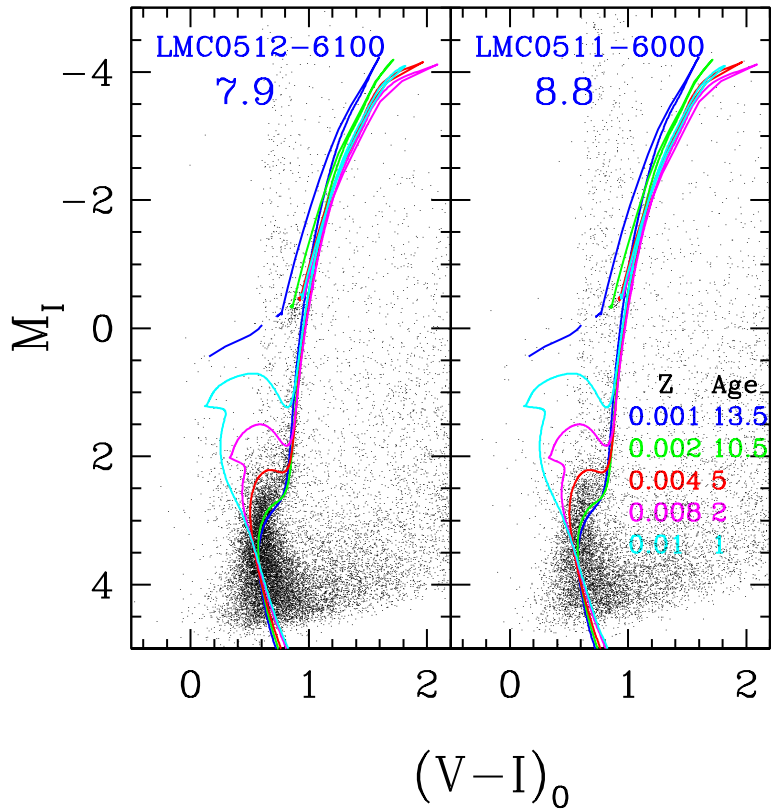}
\caption{[$(V-I)_{0}$,$M_{I}$] CMDs for the northern fields. As in Figure
\ref{fig2}, isochrones have been superimposed. The identification of each field together with its galactocentric distance, in kpc, are labeled in top-left corner.\label{fig3}}
\end{figure}

[$(B-I)_{0}$,$M_{I}$]
CMDs of the western (towards the SMC) and eastern (in the opposite
direction) fields are shown in the bottom and top panels of Figure
\ref{fig2}, respectively. [$(V-I)_{0}$, $M_{I}$] CMDs of the
northern fields are plotted in Figure \ref{fig3}. Although we postpone
to a forthcoming paper the detailed analysis of these CMDs, a set
of isochrones has been overimposed on them with the purpose of intercomparing
fields. We used the same set of overshooting isochrones as in Paper
I selected from the BaSTI library \citep{pietrinferni2004}. Their
metallicities follow approximately the AMRs derived in Paper II (see section \ref{amr}).

All fields studied, even those at a distance of $\sim$9 kpc, have a well populated old main-sequence turn-off region. This, in agreement with the results obtained in Paper I, suggests that star formation started at about the same time in all of them. An alternative is that these stars were formed in the inner disk and then they have migrated to the outskirts. All fields contain an important amount of intermediate-age LMC populations. The bulk of stars in the innermost field studied, LMC0422-7100, have ages older than 2 Gyr while most of the stars in the remaining fields have ages older than 5 Gyr. In all fields a number of stars overlapping the 1 Gyr can be observed, which number increases as the global number of the stars in the field increases. The presence of this population was also noted by \citet{gallart2004} in a field located at $\sim$7 kpc and interpreted as a burst of star formation 1 Gyr ago. However, since they are present in all studied fields and their number is correlated with the density of stars in each field they may be blue stragglers. As in the case of Paper I the comparison with isochrones show an increase of the  age of the intermediate-age populations present in each field with galactocentric distance. There are not clear differences with azimuth since CMDs of fields at similar distances but different azimuths have a similar morphology.

\section{Spectroscopic Observations, Data Reduction and Radial Velocities}\label{datareduction}

As in Paper II, the stars observed spectroscopically were selected
on the upper part of the RGB trying not to exclude metal-poor objects
on the blue ridge but avoiding inclusion of red-clump and red supergiant
objects. Stars below the tip of the RGB were given the highest priority
of observation and brighter objects were only targeted when no others
were available. The observations were carried out in service mode
on ESO period 82 with FLAMES mounted at VLT UT2. We used the low-resolution
grating LR8 which covers the wavelength range between $\sim$ 8200
to $\sim$ 9400 \AA~with a resolution of R $\sim$ 6500, in order
to accurately measure equivalent widths of the CaT lines as well as
radial velocities.

The ESO provides reduced data obtained with the GIRAFFE data-reduction
pipeline \citep{melo2009}. This pipeline performs the bias, flatfield
and scattered light corrections, find, trace and extract the spectra
and obtain the wavelength calibration based on daytime calibration
exposures. The only reduction step that this pipeline does not perform
is the sky subtraction. To do this, we followed a similar procedure
as in Paper II but with some modifications as described by \citet{battaglia2008}.
In brief, for each exposure we average the spectra of all fibers placed
on the sky to obtain a high signal-to-noise ratio (S/N) sky spectrum. This sky spectrum, and the
objects spectra are separated
into two components: continuum and line spectra. To obtain the continuum
we fitted a polynomial rejecting the contribution of both sky and
object lines. The line spectrum is obtained by subtracting the continuum.
The sky-line component is cross correlated with each object-line to
put both in the same wavelength scale. This also provides an additional
test of the wavelength calibration. After updating the wavelength
calibration, the sky- and object-line components are compared looking
for the scale factor that minimises the sky line residuals over the
whole spectral region. The object-continuum component is wavelength
updated and added back to the sky subtracted object-line spectrum.
Finally, the sky-continuum is subtracted assuming that the scale factor
is the same than for the sky-line component. The result is that the
residuals of the sky lines are lower than $\sim$3\%.

Since 2 or 3 exposures were obtained for each configuration we combined
all the spectra of the same object, thus improving the S/N and minimising
the contribution of cosmic rays and bad pixels. Finally, each object
spectrum is normalised.

Radial velocities are key to reject foreground objects, but they also
provide valuable information about the LMC kinematics (see Section
\ref{kinematics}). However, radial velocity standard stars were not
observed, so we used one of our targets with a very high S/N as template.
We obtained the radial velocity of this star using the laboratory
position of the three CaT lines and other lines easily identified
in its spectrum using the \textit{rvidlines} task in IRAF. Then, the radial velocities of all stars were calculated by cross-correlation
using the IRAF \textit{fxcor} task, with this high S/N star as template.
As in paper I, we considered as LMC members those stars with radial
velocities in the range 170$\leq$V$_{r}\leq$380 km s$^{-1}$ \citep{zhao2003}.
 In total, we have determined ages, and metallicities
for 147 stars confirmed as LMC members from their radial velocities
in the 6 fields observed. Their magnitudes, CaT equivalent widths (see below),
and radial velocities are listed in Table \ref{tbl2}.

\section{CaT Equivalent Widths and Metallicity Determination}\label{metaldetermination}

The metallicity of each star was determined from the equivalent widths
of the CaT lines using the procedure described by \citet[hereafter Paper III]{carrera2007}.
The line profiles were obtained by fitting a Gaussian plus a Lorentzian,
which provides the best fit to line core and wings (see Paper III
for details). Although the spectra had already been normalised, the
position of the continuum was recalculated by a linear fitting to
the mean values of each continuum bandpass defined in Paper III. The
CaT index, denoted as $\Sigma$Ca, is defined as the sum of the equivalent
widths of the three CaT lines.

The strength of the CaT lines depends on the temperature, gravity
and chemical abundance of the stellar atmosphere. Since we are only
interested on the abundance dependence, the temperature and gravity
contributions must be removed. The procedure followed is described
in depth in Paper III and we summarize here the most important details.
For a given chemical abundance, the stars define a sequence in the
Luminosity-$\Sigma$ Ca plane when the temperature and/or gravity
change. Although this sequence is not linear for the whole magnitude
range covered by the RGB it is very close to linear when only stars
in the upper two magnitudes of the RGB are considered as in our case.
The non-linearity of the sequences in the Luminosity-$\Sigma$ Ca
plane is in general important for [Fe/H]$\leq$-3 \citep{starkenburg2010}. However, the
expected number of stars in that metallicity range is negligible based
on the results described in Paper I. We safely assume therefore that
the sequence defined by stars of the same metallicity in the Luminosity-$\Sigma$
Ca plane is linear. We used M$_{I}$ as luminosity indicator, because
the CaT metallicity calibration based on M$_{I}$ is less sensitive
to age. In the M$_{I}$-$\Sigma$ Ca plane the sequences can be parametrised
in the form $\Sigma Ca=W'_{I}+\beta_{I}M_{I}$.
W'$_{I}$ is called the reduced equivalent width. The slope is independent
of the metallicity,
whereas W'$_{I}$ changes with the chemical abundance.

In Paper III, we obtained relationships between W'$_{I}$ and metallicity in different
metallicity scales, including the \citet[hereafter CG97]{CG97}. The
stellar metallicities of the objects studied in Paper II are calculated
in this scale. Recently \citet[hereafter C09]{carretta2009} have
updated their metallicity scale based on more homogeneous and higher
resolution spectra of a larger sample of bright giants in 24 Galactic
globular clusters. They provide the relation between their new and old scales.
The existence of this new metallicity scale for globular clusters
as well as new determinations of metallicities for some open clusters
used in Paper III from high-resolution spectroscopy (R$\sim$20000)
motivate us to obtain a new relation between the strength of the CaT
lines and metallicity. To do this we used the same cluster sample
as in Paper III and listed in Table \ref{tbl3}. The reduced equivalent
width of each cluster in $V$ and $I$ shown in columns 2 and 3 are
the same of those used in Paper III. The metallicities used in Paper
III, the new values calculated here and the corresponding references are listed
in column 4, 5 and 6, respectively.

\begin{figure}
\includegraphics[angle=0,scale=.50]{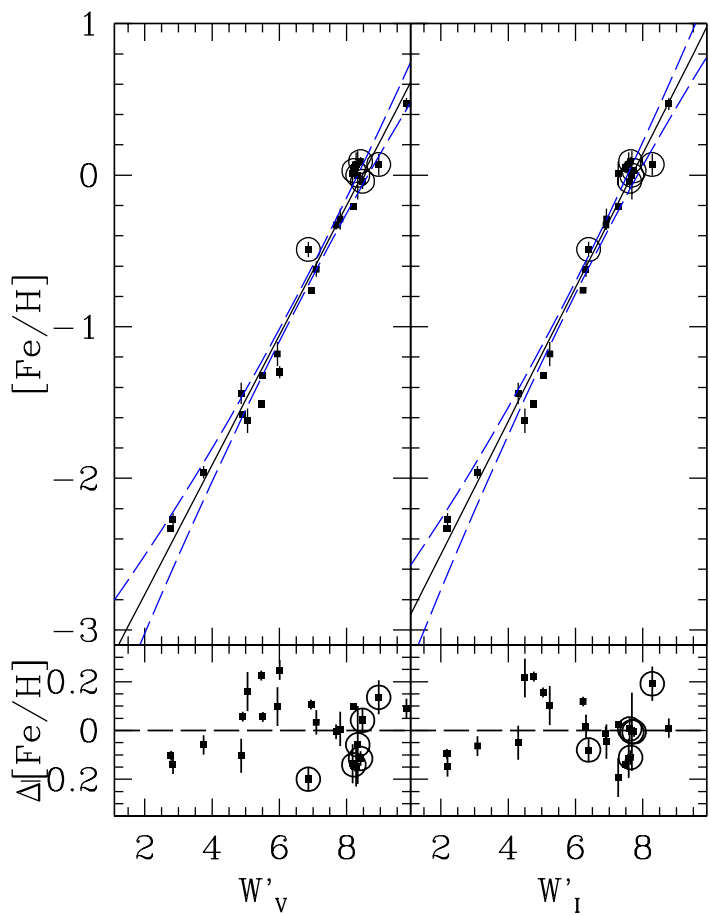}
\caption{Top panels: [Fe/H] vs. W'$_{V}$ (left) and W'$_{I}$ (right). The solid lines are the best linear fits to the data.
Dashed lines represent the 90\% confidence level band of the fits.
Open circles are clusters younger than 4 Gyr. Bottom panels: the residuals
of the linear fits. Note that the W' errors are smaller that the size
of points in several cases.\label{calCaT}}
\end{figure}

The behaviour of W'$_{V}$ and W'$_{I}$ as a function of metallicity
are shown in Figure \ref{calCaT}. We have used 25 clusters
for the calibration in $V$ and 23 for that in $I$. In both cases
a linear correlation is obtained with best fits given by:

\begin{eqnarray}
 & [Fe/H]^{V} & =-3.62(\pm0.02)+0.43(\pm0.01)W'_{V},~~~~~\sigma_{V}=0.03\\
 & [Fe/H]^{I} & =-3.38(\pm0.02)+0.44(\pm0.01)W'_{I},~~~~~\sigma_{I}=0.03\label{calCaTI}
\end{eqnarray}

In Paper III we determined that the linear sequence in the M$_{I}$-$\Sigma$
Ca plane is given by $\Sigma Ca=W'_{I}-0.611\times M_{I}$. Therefore,
using Equation \ref{calCaTI} the metallicities of the LMC stars are
given by:

\begin{equation}
[Fe/H]=-3.38+0.44\Sigma Ca+0.27M_{I}\label{eq_metal}
\end{equation}

To obtain the absolute magnitudes of the LMC stars we assumed a distance
modulus of $(m-M)_{0}=18.5\pm0.1$ and the reddenings listed in Table \ref{tbl1}
obtained from \citet{schlegel1998}.

\begin{figure}
\includegraphics[angle=0,scale=.50]{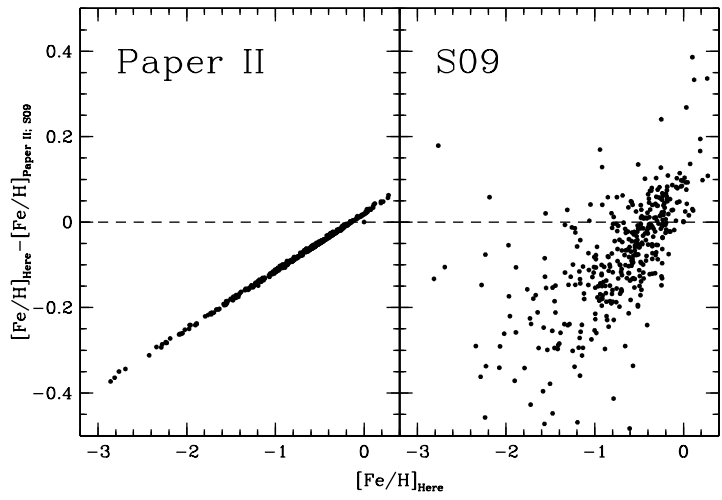}
\caption{Differences between metallicities derived here with those obtained in Paper II (left panel) and those calculated following \citet[right panel]{starkenburg2010} as a function of the values used in this paper. For this comparison we used only stars in the northern fields.\label{comparison_scales}}
\end{figure}

Since different metallicity scales are available in
the literature, it is necessary to determine to what extent the metallicity
distributions obtained here depend on the scale used. To do this we
have calculated the metallicities of all stars observed in the northern
fields ($\sim$ 400) in three different scales, which we display as
follows. The left panel of Figure~\ref{comparison_scales} shows
the difference of metallicities obtained in Paper II and those obtained here as a function of the last ones. The
new values are systematically more metal-poor than those derived in
Paper II, and the difference increases towards lower metallicities.
This is explained by the difference between the two globular cluster
metallicity scales used in each case \citep[for a detailed comparison see][]{carretta2009}.
It is interesting to note that all the CaT surveys performed until now in dwarf spheroidals companions of the Milky Way do not find objects
as metal-poor as those observed in the Galactic halo \citep[e.g.][]{helmi2006}.
The new scale seems to solve in part this discrepancy. However, \citet{starkenburg2010} demonstrated  
that the lack of metal-poor stars in CaT surveys is due to the way in which the temperature and gravity contributions are
taken into account. They proposed that extremely metal-poor stars
are not detected because of the wrong assumption that stars of the
same metallicity follow a linear sequence in the M$_{I}$-$\Sigma$
Ca plane (see also Paper III). On the right panel of Figure \ref{comparison_scales}
we have compared the difference of metallicities derived using the \citet{starkenburg2010}
recipe to those obtained here with our method. As before, our metallicities are systematically lower than
those by \citet{starkenburg2010} at low metallicities. However in this case the dispersion is much larger. We conclude that the differences
between the three scales are only noticeable ($\Delta$[Fe/H]$>$0.2 dex) for extreme values of
the metallicity but that in the metallicity range which contains most
of the LMC stars (i.e., -1.6$\leq$[Fe/H]$\leq$+0.3) they produce
values within $\pm$0.2 dex.

The metallicity distributions on each field are shown in Figures \ref{metal_norte}
and \ref{metal_otros}, and will be discussed \emph{in extenso} in
Section \ref{metaldistributions}.

\section{Estimation of Stellar Ages}\label{agedetermination}

The age-metallicity degeneracy in the position of the RGB in the CMD
can be broken when either the metallicity or the age can be obtained independently
of each other. Because of this, combining the metallicities of stars
derived from spectroscopy with their position in the CMD allows us
to estimate their age. In Paper II, a polynomial relationship was
computed to derive stellar ages from their metallicities and positions
in the CMD. For this purpose, a synthetic CMD was created with IAC-star
\citep{aparicio2004}, adopting BaSTI stellar evolution models with
overshooting \citep{pietrinferni2004}. The aforementioned relation
was obtained in Paper II for ($V-I$) and M$_{V}$ while \citet{carrera2008b}
derived a similar relation from the same synthetic CMD for ($B-I$)
and M$_{I}$. We refer the reader to these papers for a detailed discussion
of the procedure used, uncertainties and tests of confidence levels.
\citet{carrera2008b} confirmed that both relationships produce similar
results within the uncertainties. We have used the corresponding coefficients -those derived for ($V-I$)
in the case of the North fields, and those for ($B-I$) for the remaining
fields- obtained in these papers to derive ages for each program star.

\section{Analysis}

\subsection{Metallicity Distributions}\label{metaldistributions}

\begin{figure}
\includegraphics[angle=0,scale=.50]{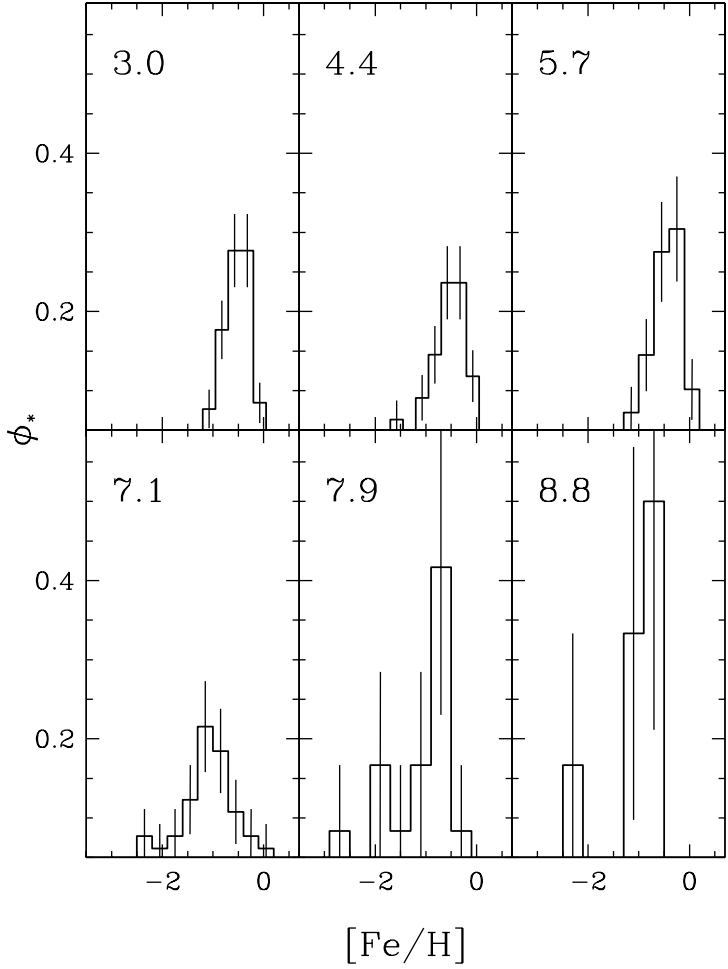}
\caption{Metallicity distributions for the northern fields. The innermost field
is on the top-left and the outermost one on the bottom-right. The
galactocentric distances in kpc of each field are labeled on the
top-left corner. The four innermost fields were the subject of Paper
II, for which metallicities have been recalculated in the new scale
defined in Section~\ref{metaldetermination}. The errorbars are the square root of the number of stars in each bin.\label{metal_norte}}
\end{figure}

\begin{figure}
\includegraphics[angle=0,scale=.50]{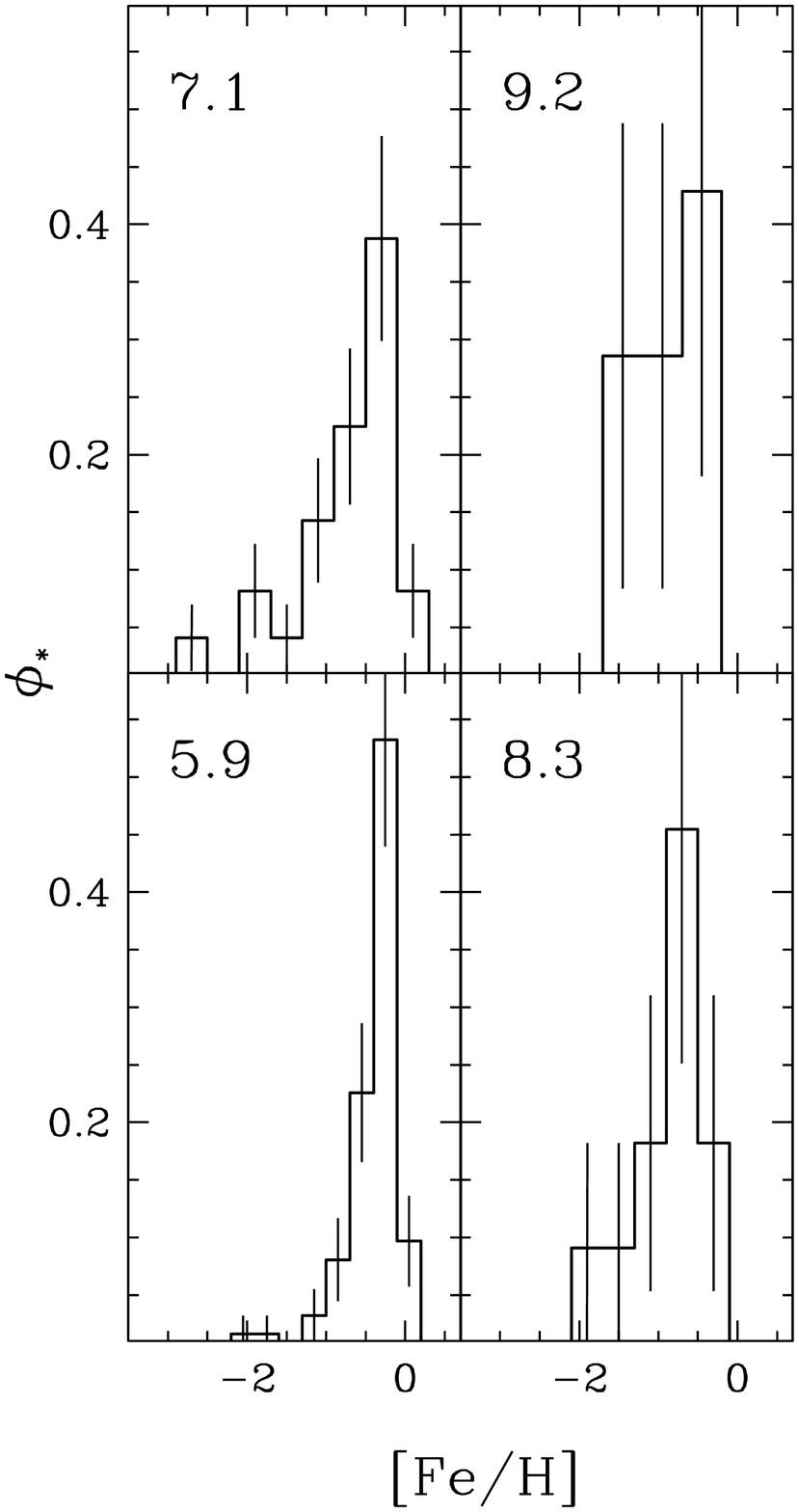}
\caption{The same as Figure \ref{metal_norte}, for the eastern (top) and western
(bottom) fields.\label{metal_otros}}
\end{figure}

The metallicity distributions for each field are shown in Figures~\ref{metal_norte}
and \ref{metal_otros}. Because we used here a different metallicity
scale than in Paper II, we have also recalculated the metallicity
distributions of the fields analysed there (Figure \ref{metal_norte}).
The differences between the distributions described in Paper II and
those obtained with the new scale are only appreciable for low metallicity
stars, which are poorly sampled. The small number of objects ($<50$)
confirmed as LMC members from their radial velocity in the new fields
studied, except for field LMC0422-7100 (see Table \ref{tbl5}), do
not allow us to perform a detailed analysis of these distributions.
However, we can use their mean metallicity to investigate possible
differences among them as well as metallicity gradients. In Paper
II means and dispersions were obtained by Gaussian fitting to each
distribution. This is correct for samples with a number of objects
large enough ($\gtrsim$50). But Gaussian fitting is not adequate for
poorly sampled distributions. To overcome this problem we have adopted
the following procedure. Half of the stars in each sample were randomly
selected and the mean, standard deviation and 10th and 90th percentile
of their metallicity distribution were computed. This procedure was
repeated 10$^{4}$ times using random subsets each time. The values
obtained in all of them for each statistical parameter have Gaussian
distributions centered in the true mean, dispersion or percentile
value. The width of each distribution provides an estimation of its
uncertainty. The values thus obtained are listed in Table \ref{tbl5},
where the fields are ordered by their distance to the center (see
Section \ref{spatial_distribution}) given in column 2. Uncertainties
are smaller for better sampled fields, as expected. The RGB sample
studied in the LMC bar by \citet{cole2005} has been also analysed
in the same way and the results are also listed in Table \ref{tbl5}.
Since their CaT index is equivalent to ours, as was demonstrated in
Paper III, we used their index to calculate the metallicities from
Equation \ref{eq_metal}.

\begin{figure}
\includegraphics[angle=0,scale=.50]{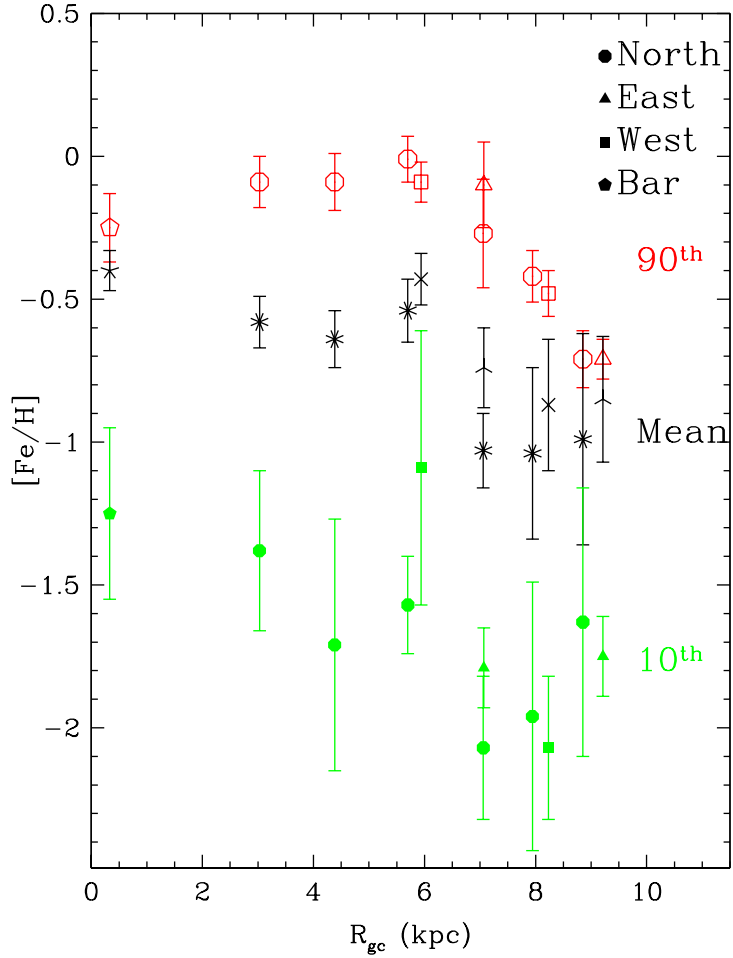}
\caption{Metallicity at the 10th (metal-poor; green filled symbols) and 90th (metal-rich; red
empty symbols) percentile and mean (black starred symbols) values of the metallicity distribution of
each field as a function of projected radius. See text for details.\label{percentiles}}
\end{figure}

The mean metallicity values in the inner $\sim$7 kpc are similar
within the uncertainties but are lower than $\sim$-0.8 dex beyond this radius
suggesting the presence of a gradient. The computed 10th and 90th
percentile values provide further insight into the possible presence
of a gradient. They are listed in columns 5 and 6 of Table \ref{tbl5},
and have been plotted in Figure \ref{percentiles} as a function of
radius. The 10th percentile value (filled symbols), account for the
most metal-poor regime and have similar values in all disk fields
and in the bar within the uncertainties. The 90th percentile values
(open symbols) which account for the most metal-rich populations,
are however, constant within the inner $\sim$7 kpc, but decrease
from this point on, indicating a metallicity gradient in the most
metal-rich populations outwards of $\sim$7 kpc. The implications
of this gradient are discussed in depth in Section \ref{sec_discuss}.

\subsection{Age-Metallicity Relationships}\label{amr}

\begin{figure}
\includegraphics[angle=0,scale=.50]{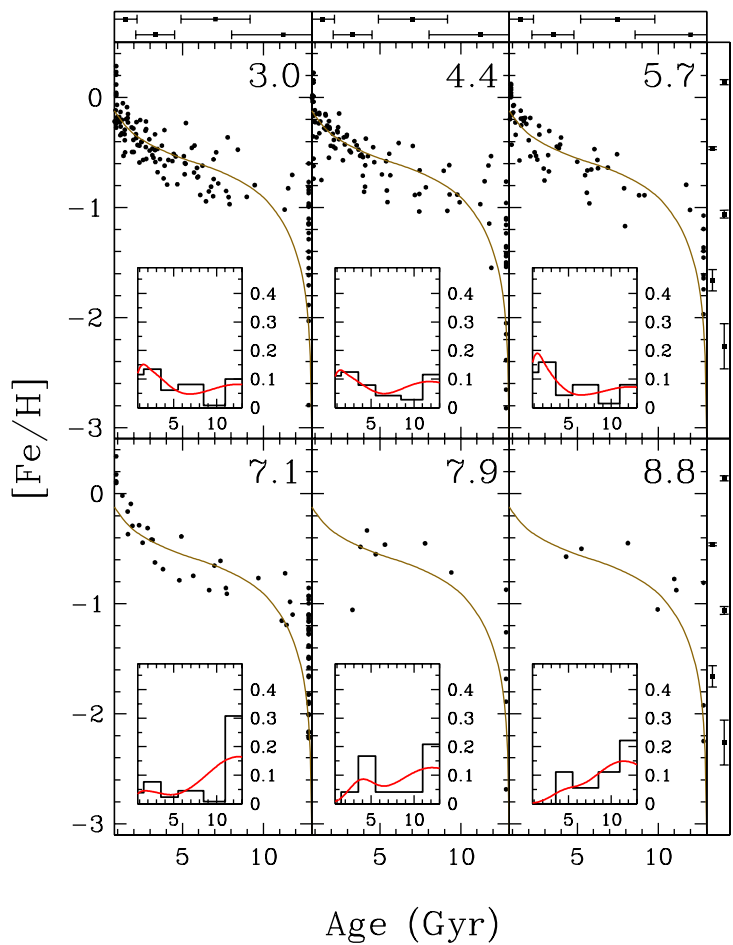}
\caption{Age-metallicity relationships for the northern fields in our sample.
One of the chemical enrichment models derived in Paper II has been
overplotted for comparison (see text for details). The galactocentric
distances of each field are labeled in top-right corner, in kpc.
Inset panels show the age distribution computed by taking (solid line)
and by not taking (histogram) into account the age determination uncertainties
(see Paper II for details). Age error are given in the top strip while
metallicity errors are shown in the right one.\label{amr_norte}}
\end{figure}

\begin{figure}
\includegraphics[angle=0,scale=.50]{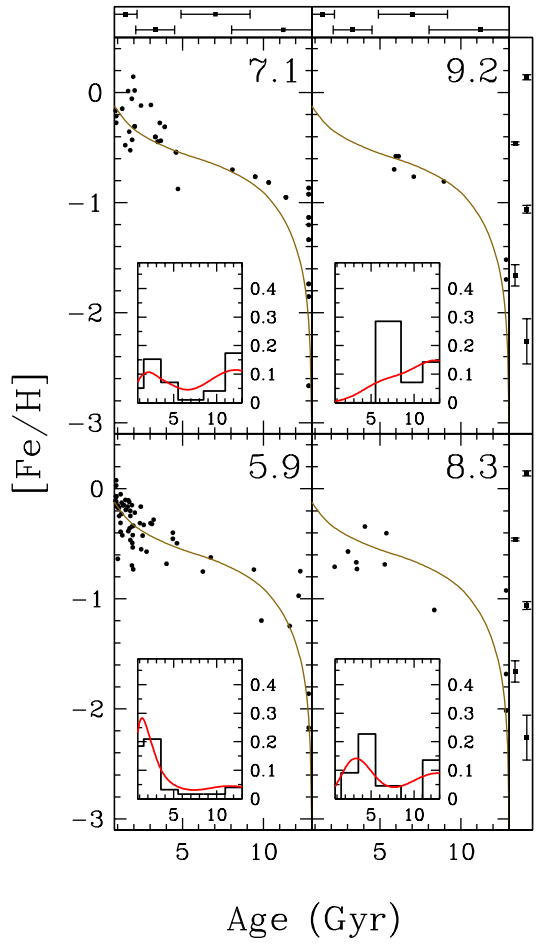}
\caption{The same as Figure \ref{amr_norte}, for eastern (top) and western
(bottom) fields in our sample.\label{amr_otros}}
\end{figure}

We will further investigate the nature of the changes in the spatial
metallicity distribution with the help of the AMRs of each field.
In spite of the large uncertainties of the individual age determinations
presented in Section~\ref{agedetermination}, the AMRs are useful
to characterise the general trend of the run of metallicity with age.
The AMR for each field is plotted in Figures \ref{amr_norte} and
\ref{amr_otros}. The many stars with the same age that appear both
at the old and young limits result from the method degeneracy. As
discussed in Paper II and in \citet{carrera2008b}, our procedure to
estimate ages saturates for values older than 10 Gyr, so we can only
give 10 Gyr as a lower limit for older stars. We assume for them an
age of 12.9 Gyr, the age of the oldest globular cluster observed in
our galaxy \citep[NGC~6426, ]{salaris2002}. In the case of the youngest
stars in our sample, stellar evolution models indicate that the region
in which we selected the observed stars is not populated by objects
younger than 0.8 Gyr. Therefore, although equations used to estimate ages can
formally lead to ages younger than this value, we assign them an age
of 0.8 Gyr.

All fields studied, independently of their distance to the center,
show a rapid chemical enrichment reaching [Fe/H]$>$-1 in the
first 2--3 Gyr. After this, the metallicity increased steadily but
moderately until around 3 Gyr ago, thus increasing another $\sim$0.5
dex in about 7 Gyr. This period corresponds with the age-gap observed
in the clusters \citep[e.g.][]{grocholski2006,sharma2010} and with
a period of lower star formation rate observed in the field \citep[e.g.][]{meschin2010}.
Finally, another episode of rapid chemical enrichment in the last
$\sim$3 Gyr is only observed in fields inner to $\sim$7 kpc. In
fact, we do not find stars younger than $\sim$3 Gyr in our sample
in the outermost fields despite the fact that these are among the
most common on the innermost fields. The two periods of chemical enrichment
at old and young ages are also observed in the clusters \citep[e.g.][]{olszewski1991}.
We refer the reader to Paper II for a detailed comparison among clusters,
bar, and disk AMRs. In Paper II, we performed a $\chi^{2}$ test to
check that the AMRs of the fields studied there are statistically
indistinguishable among them. Due to the small number of stars confirmed
as LMC member in the outermost fields studied here, a $\chi^{2}$
test is not adequate in this case and we will restrict ourselves to
a qualitative comparison among the AMRs of the different fields. To
help with this comparison, we have plotted in Figures \ref{amr_norte}
and \ref{amr_otros} one of the chemical evolution models derived in
Paper II that reproduces adequately the observed AMRs. This model was
computed assuming a rate of gas outflow of 2.5 (see Paper II for details)\footnote{This model has been recomputed with the metallicity scale used in
this paper.%
}. The AMRs of all fields are well reproduced by this model and therefore
we feel confident that the chemical enrichment has proceded in a similar
way in all the studied fields.

Inset panels of Figures \ref{amr_norte} and \ref{amr_otros} show
the age distribution of the stars measured on each field, with and
without taking into account the age uncertainty (solid line and histogram,
respectively; see Paper II for details). In the innermost fields there
are two peaks which correspond to the two periods of faster chemical
enrichment. The amount of stars younger than 5 Gyr gradually decreases
with increasing radius. Since the younger stars are also more metal-rich,
the metallicity gradient observed in the outer disk may be explained by
the gradual increase of the age of the youngest populations with radius.

\subsection{Kinematics of the LMC RGBs}\label{kinematics}

The kinematics of the LMC has been studied by many authors using different
tracers \citep[see][for a review]{vandermarel2009}. There is agreement
among workers that all tracers studied describe a similar cold disk
kinematics but with different rotational velocities for each of them: 61 km s$^{-1}$
for the carbon stars \citep{vandermarel2002}; 80 km s$^{-1}$ for
\ion{H}{1} gas \citep{olsen2007}; and 107 km s$^{-1}$ for red
supergiants \citep{olsen2007}. It is also known that the velocity
dispersion, related to the disk scale height, increases from the youngest
\citep[red supergiants $\sim$9 km s$^{-1}$,]{olsen2007} to the oldest
populations \citep[RR Lyrae stars $\sim$50 km s$^{-1}$,]{borissova2006,minniti2003}.
The fact that the oldest populations have the highest velocity dispersion
suggests the possibility that they may be distributed in a halo similar
to that of the MW \citep[e.g.][]{minniti2003,borissova2006}

\begin{figure}
\includegraphics[angle=0,scale=.50]{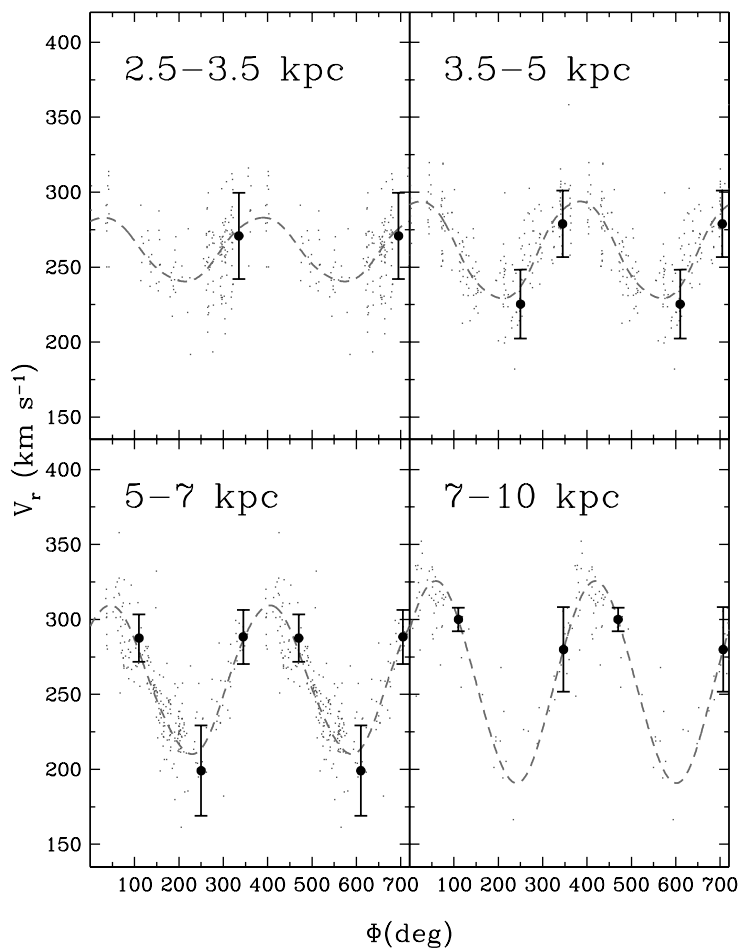}
\caption{Mean and dispersion of the velocity distributions (large dots with
errorbars) as a function of the position angle (note that each distribution
is plotted twice) at different angular distances (from top-left to
bottom-right). As comparison, carbon stars data have been overplotted
\citep[small light grey dots, from ][]{kunkel1997,hardy2010}. Dashed
lines are the predictions of the kinematical model computed by \citet{vandermarel2002}
for carbon stars.\label{fig_rotation}}
\end{figure}

Our sample of almost 500 RGB stars studied in an homogeneous way covers
a large range of angular distances. Larger samples have been studied
but always at lower galactocentric radius \citep[e.g.][]{zhao2003,cole2005}.
In Figure~\ref{fig_rotation} we have plotted the velocities of the
RGB stars studied here versus position angle for different galactocentric
radius (filled circles). Our RGB sample is well reproduced by the
kinematical rotational disk model derived by \citet{vandermarel2002}
for carbon stars (dashed line). In Figure \ref{fig_rotation} the
carbon stars data (dots) from \citet{kunkel1997} and \citet{hardy2010}
have also been plotted as comparison. The velocity dispersion of the
RGB sample studied here ($\sigma$=24.1$\pm$0.2 km s$^{-1}$) is
slightly larger than that of the carbon stars ($\sigma$=20.2$\pm$0.5
km s$^{-1}$). Our result is in very good agreement with other analysis
of LMC field RGB by \citet[$\sigma$=24 km s$^{-1}$,]{zhao2003} and
\citet[$\sigma$=24.7$\pm$0.4 km s$^{-1}$,]{cole2005}.

Since RGBs have a wide range of metallicities and ages, we can compare
the velocity dispersion of different ages/metallicities, which provides
information about the scale height of different populations. In fact,
if a classical halo formed by old, metal-poor, high-velocity dispersion
objects existed in the LMC, a dependence between the velocity dispersion
and metallicity might be observed. To investigate this we have compared
the velocity dispersion in different metallicity bins. The mean velocity
of each field was subtracted from the radial velocity of each star
to eliminate the rotational component, as in Paper II. Then, we divided
the metallicity range in 4 bins and computed the velocity dispersion of each
of them by fitting a Gaussian to the velocity distribution. The number of objects and velocity dispersion in each metallicity bin are listed
in Table \ref{tbl6}. It can be noted that the velocity dispersion
increases as metallicity decreases, or age increases. This agrees
with the fact found by other authors that the velocity dispersion
increases from the youngest population to the oldest ones \citep[e.g.][]{olsen2007,vandermarel2009}.
The velocity dispersion of our most metal-poor, and therefore older,
bin is $\sigma$=26.8$\pm$1.4 km s$^{-1}$ based on 45 objects. This
value, similar to that found in Paper II from a smaller sample, is
lower than the dispersion found by \citet{cole2005} for the most
metal-poor RGBs in the bar, $\sigma$=40.8$\pm$1.7 km s$^{-1}$ based
only on 18 stars. Although we believe that we are sampling the most metal-poor/oldest populations, we  do not find evidences that they belong to a hot halo. \citet{borissova2006} \citep[see also][]{minniti2003}
found $\sigma$=50$\pm$2 km s$^{-1}$ for a sample of RR Lyrae in
the center of the LMC. These values are significantly higher than the value found here
for the most metal-poor stars. These larger velocity dispersions have
been quoted as evidence for the presence of a classical halo.
However, this is still an open question since most RR Lyrae have an exponential
surface density distribution with a similar scale length as other
well known tracers of the disk \citep{alves2004,subramaniam2009}, and only a small fraction, $\leq$10\%, is likely to trace the extended halo of the LMC \citep{subramaniam2009}.

\begin{figure}
\includegraphics[angle=0,scale=.50]{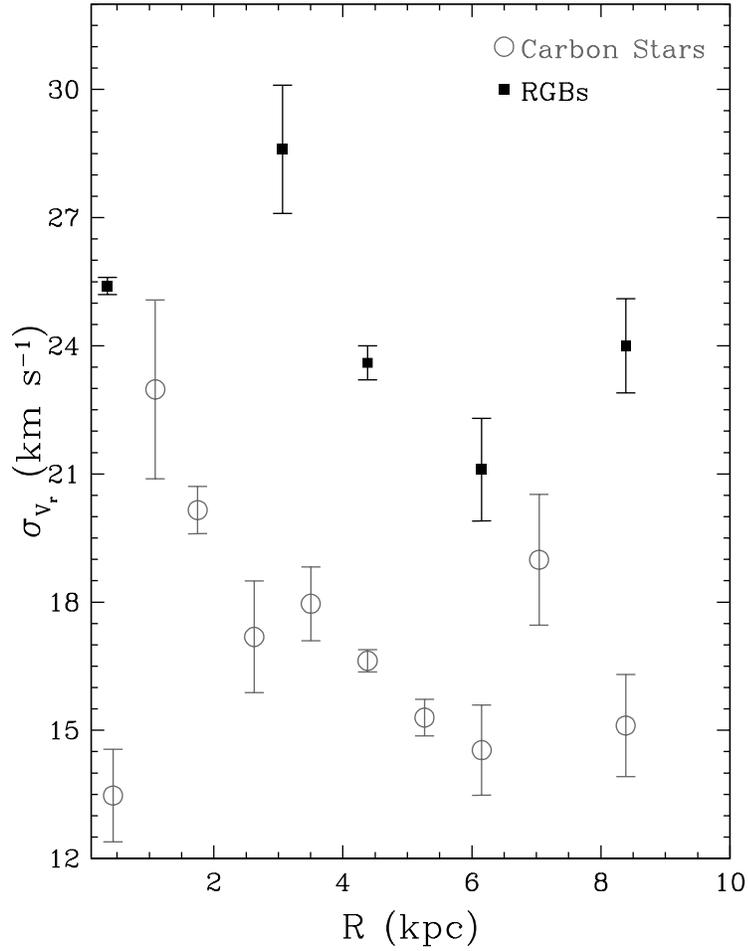}
\caption{Radial velocity dispersion as a function of radius for the RGBs (filled
squares) and for the carbon stars (open circles). With the exception
of the bar region, both show a decrease of the velocity dispersion
with radius up to a distance of 6 kpc. From there off, an increase of velocity dispersion is observed although the small number of objects studied at these galactocentric distances do not allow us a reliable discussion.\label{fig_vel_dispersion}}
\end{figure}

The galactocentric distance covered by our sample also allows us to
investigate the behaviour of the velocity dispersion as a function
of radius. This is a key analysis to investigate the stability of
the disk and, to our knowledge, this is the first time it is done
in the LMC. As before, the velocity dispersion in four radial annuli
has been computed by fitting a Gaussian to the rotational-free velocities
of stars in each of them. The obtained values, together with the number
of objects in each annulus, are listed in Table \ref{tbl7} and plotted
as filled squares in Figure \ref{fig_vel_dispersion}. The velocity
dispersion decreases with radius with the exception of the outermost
annulus studied. As comparison, we obtained in the same way the velocity
dispersion of the Bar from the sample presented by \citet{cole2005}.
The Bar has a smaller dispersion than the innermost disk annulus analysed.
Since our sample is not well distributed in angular distances, we
have also calculated the velocity dispersion as a function of radius
for the carbon stars sample described above (open circles). As in
the case of the RGBs, the velocity dispersion decreases with radius
if we exclude the central area. The behaviour in the outermost annulus
studied may not be reliable because of the small number of
objects studied at this galactocentric distance. The implications
of these results are discussed in depth in the following section.

\section{Discussion}\label{sec_discuss}

\subsection{Age and Metallicity Gradients in the LMC}

The spatial distribution of our fields allows us to investigate the presence
of age and/or metallicity gradients in the field populations of the
LMC. In the 4 fields studied in Paper I, located at galactocentric
distances from 3 to 7 kpc, we found that while the oldest populations
are coeval in all fields, the age of the youngest component gradually
increases with radius. With the new data, we confirm that the trend continues at larger galactocentric distances ($\sim$9 kpc). If we assume
that the youngest stars were formed in-situ (e.g. there is no significant
stellar migration), we are observing an outside-in quenching of star
formation at recent times.

This age gradient observed across the LMC disk is not accompanied by
a gradient in the mean metallicity, which it is observed only for R$\gtrsim$6
kpc. A gradient in the metallicity in the youngest objects cannot be excluded with our observations since they do not have counterparts in the red giant branch. However, the fact that the AMR is so similar at all radii points
to a similar history of metal enrichment across the whole LMC disk.
In this context, a gradient in the mean metallicity of different fields
would be mainly related to a varying proportion of stars of different
ages in different fields. A similar trend of metallicity with radius has been found by \citet{feast2010} for AGB stars \citep[but see also][]{cioni2009}.
Recent studies in clusters also show hints of a possible metallicity
gradient in the outer disk \citep[Figure 16 of][]{sharma2010}. However,
the small number of clusters at these outer distances prevents them from
extracting further conclusions.

A dynamical bar, as that observed in the LMC, might be able to mix
the stellar content of a disk within a certain radius \citep{friedli1994,zaritsky1994}.
Therefore this result might put a constrain to the mixing produced by the
LMC bar. The age gradient observed at R$<$6 kpc can be explained
by the fact that the galaxy has not had enough time to mix populations
younger than 1 Gyr. It is expected that the stellar content of the
LMC is well mixed in a few crossing times. At a distance of 6 kpc, the crossing time\footnote{This value has been computed from the  $t_{cross}=R/\sigma$ \citep[e.g.][]{galacticdynamics1987} using a velocity dispersion of $\sigma$=24.1 km s$^{-1}$ derived in previous section.} is $\sim$2$\times10^{8}$ yr which is in agreement with what observed.

The mean metallicity of the most metal-rich objects in each field gradually
decreases with the galactocentric distance from R$\sim$6 kpc. This trend is due to an increase of the age of the youngest population
in each field with radius, because the chemical enrichment history has been
the same over the LMC disk. A similar trend has been also observed in the M33 outer disk \citep{barker2007}. \citet{taylor2005} studied the behaviour of color with radius in a sample of irregular and late-type spiral galaxies finding that these galaxies became redder with increasing radius which is also interpreted as the existence of an age gradient. This agrees
with an outside-in disk evolution scenario proposed in Paper I for
the LMC as opposed to the inside-out scenario predicted by $\Lambda$CDM
cosmology \citep{white1991,mo1998} and observed in more massive early-type
spirals \citep[e.g.][]{trujillo2005}. This outside-in disk evolution
scenario is qualitatively similar to that predicted by \citet{stinson2009}
for dwarf galaxies and observed in several Local Group dwarfs in the
LCID project \citep[e.g.][]{hidalgo2009}. However, for a galaxy like
the LMC, the \citet{stinson2009} models predict an inside-out scenario.
This may put constraints to the mechanisms that shape the radial distribution
of stellar populations in galaxies of intermediate mass such as the
LMC.

It is well known that the LMC forms an interacting system with the
SMC and the MW at the moment. However it is not clear if the same
has happened in the past. New dynamical models, based on accurate
proper motions \citep{kallivayalil2006,piatek2008}, suggest that the Magellanic Clouds are unbound and
they might be in their first passage around the Milky Way \citep{besla2007}.
Repeated encounters with one or more external systems
might have erased, or at least weakened, any gradient. The
results of this paper tend to support an evolution of the LMC in a more isolated
environment.

\subsection{Kinematics of the RGB Field Populations}

Our RGB sample has rotational cold disk kinematics similar to that
of carbon stars. The velocity dispersion is $\sigma$=24.1$\pm$0.2
km s$^{-1}$. It increases from most metal-rich/youngest stars, $\sigma$=20.9$\pm$1.2
km s$^{-1}$, to the most metal-poor/oldest ones, $\sigma$=26.8$\pm$1.4
km s$^{-1}$. Other tracers have a similar behaviour \citep[see][for a review]{vandermarel2009}.
Therefore, the scale height of the LMC disk increases with time. However,
even for the most metal-poor/oldest objects there are not evidences
of the presence of a classical hot halo. In fact, recent works indicate
that the LMC disk extends, at least, to a distance of 12-14 kpc \citep{majewski2009,saha2010}.

The velocity dispersion of disk RGBs decreases with increasing galactocentric
radius, except for those objects in the bar region, which have a lower
velocity dispersion than those in the inner disk. Carbon stars
have a similar trend. From pioneering works in the disk of spiral
galaxies it is well known that the velocity dispersion decreases with
increasing radius \citep[e.g.][]{bottema1993}. The same trend has been
also found in the only two disk irregular galaxies studied until now
to our knowledge: NGC~2552 \citep{swaters1999}; and NGC~4449 \citep{hunter2005}.
This trend is easily explained by the decreases of stellar density
with galactocentric distance if the scale height remains constant
with radius, as is observed in most disk galaxies \citep{vanderkruit1981}.
The fact that the velocity dispersion is lower in the bar than in
the innermost disk is a feature also observed in at least one half
of barred spirals \citep[e.g.][]{chung2004}. The velocity dispersion
increase observed in the outermost radius studied can be interpreted
as a thickening in the external disk due to dynamical interactions
with the MW \citep{weinberg2000}. However the small number of objects
confirmed as LMC members as this galactocentric distance prevents
us from extracting further conclusions.

\section{Summary}\label{sec_summa}

In this paper we have presented deep CMDs for 6 fields in the outer
LMC disk, covering different distances, 5.9 kpc$\leq$R$\leq$9.2 kpc,
and azimuths. Using CaT spectroscopy we have derived stellar metallicities
for a sample of RGB stars in these fields, which have been analysed
together with those stars studied in Paper II. In total, including
those studied in Paper II, our sample has 497 disk RGB stars confirmed
as LMC members from their radial velocity, covering a galactocentric
radius from 3.0 to 9.5 kpc and different position angles. The main
results of this paper are:
\begin{itemize}
\item The comparison of the CMDs with isochrones shows that the age of the
youngest populations gradually increases with radius. This is a continuation of the trend found
in Paper I for inner fields. 
\item The small number of stars confirmed as LMC members in the outermost
fields (R$>$7 kpc) does not allow a detailed analysis of the metallicity
distributions of these fields. A statistical analysis shows that the
10th percentile of the metallicity distribution, which accounts for
the most metal-poor populations, does not show a clear trend with
radius within the errors. Therefore, we may conclude that the most
metal-poor, and old, population is similar in all of them. On the
contrary, the 90th percentile, indicative of the most metal--rich
populations, remains constant in the inner 6 kpc, and gradually decreases
from that point on. This result, which is independent of the position
angle, may be explained by the gradual increase of the age of the youngest
populations with galactocentric distance. 
\item As in Paper II, we have estimated the ages of the observed stars from
the combination of spectroscopic metallicities, and positions of the
stars in the CMD. This allows us to obtain the AMR for each field.
It is consistent with a chemical enrichment that has been similar in all
studied fields as was found in Paper II for inner fields: i.e. coeval stars
would have the same metallicity everywhere in the LMC disk. 
\item Our RGB sample has a rotational cold disk kinematics similar to that
inferred from carbon stars, with a velocity dispersion of $\sigma$=24.1$\pm$0.2
km s$^{-1}$. The velocity dispersion increases from most metal-rich/youngest
stars, $\sigma$=20.9$\pm$1.2 km s$^{-1}$, to the most metal-poor/oldest
ones, $\sigma$=26.8$\pm$1.4 km s$^{-1}$ as has been observed for
other kinematical tracers \citep[see][for a review]{vandermarel2009}.
However, even for the most metal-poor/oldest objects there is not
evidence of the presence of a classical hot halo. 
\end{itemize}
\acknowledgments

We thank the referee, Dr Feast, for a useful report that pointed out a few unclear points in the original version. We also thank to I. P\`erez for her very useful comments about the literature of disks in external galaxies. A. A., C. G. and R. C acknowledge the support from the Spanish Ministry
of Science and Technology (Plan Nacional de Investigaci\'on Cient\'{\i}fica,
Desarrollo, e Investigaci\'on Tecnol\'{\i}gica, AYA2007-3E3507). R. C. also
acknowledges the funds by the Spanish Ministry of Science and Innovation
under the Juan de la Cierva fellowship. This work has made use of
the IAC-STAR Synthetic CMD computation code. IAC-STAR is supported
and maintained by the computer division of the Instituto de Astrof\'{\i}sica
de Canarias.



\textit{Facilities:} \facility{ESO 2.2m (WFI)}\facility{VLT (FLAMES)}

\clearpage

\clearpage

\clearpage

\clearpage

\clearpage

\begin{deluxetable}{lccccccc}
\tablecaption{Observed Fields\label{tbl1}}
\tablewidth{0pt}
\tablehead{
\colhead{ID} & \colhead{$\alpha_{2000}$} & \colhead{$\delta_{2000}$} & \colhead{$R_{gc}$\tablenotemark{a}} & \colhead{P. A.\tablenotemark{a}} & \colhead{E(B-V)} & \colhead{$R_{proj}$\tablenotemark{b}} & \colhead{$R (kpc)$\tablenotemark{c}}
}
\startdata
LMC0512-6648 & 05$^h$12$^m$ & -66\arcdeg48\arcmin &  3\fdg4 & 333\fdg0 & 0.10 & 3\fdg5 & 3.0\\
LMC0514-6503 & 05$^h$14$^m$ & -65\arcdeg03\arcmin &  5\fdg0 & 343\fdg3 & 0.05 & 5\fdg0 & 4.4\\
LMC0513-6333 & 05$^h$13$^m$ & -63\arcdeg33\arcmin &  6\fdg5 & 345\fdg4 & 0.04 & 6\fdg5 & 5.7 \\
LMC0422-7100 & 04$^h$22$^m$ & -71\arcdeg00\arcmin &  5\fdg6 & 250\fdg7 & 0.08 & 6\fdg8 &5.9 \\
LMC0513-6159 & 05$^h$13$^m$ & -61\arcdeg59\arcmin &  8\fdg0 & 347\fdg6 & 0.03 & 8\fdg0 & 7.1\\
LMC0649-7110 & 06$^h$49$^m$ & -71\arcdeg10\arcmin &  6\fdg9 & 110\fdg3 & 0.13 & 8\fdg0 &7.1 \\
LMC0512-6100 & 05$^h$12$^m$ & -61\arcdeg00\arcmin &  9\fdg0 & 347\fdg8 & 0.03 & 9\fdg0 & 7.9\\
LMC0356-7100 & 03$^h$56$^m$ & -71\arcdeg00\arcmin &  7\fdg7 & 250\fdg9 & 0.09 & 9\fdg3 & 8.3\\
LMC0511-6000 & 05$^h$11$^m$ & -60\arcdeg00\arcmin & 10\fdg0 & 348\fdg0 & 0.03 & 10\fdg0 & 8.8\\
LMC0714-7100 & 07$^h$14$^m$ & -71\arcdeg00\arcmin &  8\fdg9 & 109\fdg7 & 0.19 & 10\fdg4 & 9.2\\
\enddata
\tablenotetext{a}{Angular coordinates over the celestial sphere (see text for details).}
\tablenotetext{b}{De-projected radius calculated assuming $i$=34\fdg7 and $\Theta$=189\fdg3.}
\tablenotetext{b}{Distance in kpc calculated as $R[kpc]=D\times\tan(d[\arcdeg])$ assuming D=51 kpc.}
\end{deluxetable}

\clearpage

\begin{deluxetable}{lcccccccr}
\tablecaption{LMC RGB Stars Observed. Only a portion of this table is shown here to demonstrate its form and content. A machine-readable version of the full table is available.\label{tbl2}}
\tablewidth{0pt}
\tablehead{
\colhead{$\alpha_{2000}$} & \colhead{$\delta_{2000}$} & \colhead{$\Sigma$Ca} & \colhead{$\sigma_{\sigma Ca}$} & \colhead{B or V} & \colhead{I} & \colhead{V$_r$ (km s$^{-1}$)} & \colhead{$\sigma_{V_r}$ (km s$^{-1}$)} & \colhead{Comments}
}
\startdata
  04:23:02.04 & -70:58:51.0 & 8.218 & 0.262 & 19.9325 & 17.9836 & 239.6648 &  2.880 & Low S/N \\
  04:22:52.95 & -70:57:00.3 & 8.605 & 0.147 & 19.4462 & 16.8805 & 251.9283 &  1.260 & \\
  04:23:25.27 & -70:56:16.2 & 7.839 & 0.120 & 18.9702 & 16.4791 & 241.2483 &  1.494 & \\
  04:22:21.24 & -70:55:42.2 & 7.565 & 0.203 & 19.8558 & 17.7160 & 225.6490 &  2.196 & \\
  04:23:00.06 & -70:54:18.2 & 8.422 & 0.102 & 19.3202 & 16.9670 & 214.9008 &  1.587 & \\
\enddata

\end{deluxetable}

\begin{deluxetable}{lccccc}
\tablecaption{Cluster Sample\label{tbl3}}
\tablewidth{0pt}
\tablehead{
\colhead{Cluster} & \colhead{W'$_V$} & \colhead{W'$_I$} & \colhead{[Fe/H]$_{CG97}$} & \colhead{[Fe/H]$_{C09}$} & \colhead{Reference}
}
\startdata
NGC 104 & 6.94$\pm$0.01 & 6.23$\pm$0.01 & -0.67$\pm$0.05 & -0.76$\pm$0.02 & 1\\
NGC 188 & 8.17$\pm$0.07& 7.27$\pm$0.08 & -0.07$\pm$0.04 &+0.01$\pm$0.08 & 2 \\
NGC 288 & 5.51$\pm$0.01 & 5.04$\pm$0.01 & -1.07$\pm$0.03 & -1.32$\pm$0.02 & 1\\
NGC 362 & 6.01$\pm$0.01 & \nodata & -1.09$\pm$0.03 & -1.30$\pm$0.04 & 1\\
NGC 1851 & 5.94$\pm$0.03 & 5.24$\pm$0.04 & -1.03$\pm$0.06 & -1.18$\pm$0.08 & 1\\
NGC 1904 & 4.91$\pm$0.03 & \nodata & -1.37$\pm$0.05 & -1.58$\pm$0.02 & 1\\
Be 20 & 6.86$\pm$0.03 & 6.39$\pm$0.03 & -0.49$\pm$0.05 & -0.49$\pm$0.05 & 3 \\
NGC 2141 & 8.33$\pm$0.01 & 7.67$\pm$0.02 & -0.14$\pm$0.05 & +0.00$\pm$0.16 & 4 \\
Cr 110 & 8.21$\pm$0.04 & 7.74$\pm$0.06 & \nodata & +0.03$\pm$0.02 & 5 \\
NGC 2298 & 3.75$\pm$0.03 & 3.09$\pm$0.03 & -1.74$\pm$0.04 & -1.96$\pm$0.04 & 1 \\
Melote 66 & 7.69$\pm$0.03 & 6.90$\pm$0.03 & -0.38$\pm$0.06 & -0.33$\pm$0.03 & 3\\
Be 39 & 8.21$\pm$0.04 & 7.27$\pm$0.06 & \nodata & -0.21$\pm$0.01 & 6\\
NGC 2682 & 8.24$\pm$0.01 & 7.48$\pm$0.01 & -0.03$\pm$0.03 & +0.05$\pm$0.02 & 5\\
NGC 3201 & 5.46$\pm$0.03 & 4.76$\pm$0.02 & -1.23$\pm$0.05 & -1.51$\pm$0.02 & 1 \\
NGC 4590 & 2.84$\pm$0.02 & 2.19$\pm$0.06 & -1.99$\pm$0.03 & -2.27$\pm$0.04 & 1 \\
NGC 5927 &7.81$\pm$0.01 & 6.92$\pm$0.01 & \nodata & -0.29$\pm$0.07 & 1 \\
NGC 6352 & 7.10$\pm$0.01 & 6.31$\pm$0.01 & -0.64$\pm$0.06 & -0.62$\pm$0.05 & 1\\
NGC 6528 & 8.28$\pm$0.04 & 7.58$\pm$0.04 & -0.17$\pm$0.02 & +0.07$\pm$0.08 & 1 \\
NGC 6681 & 5.05$\pm$0.03 & 4.49$\pm$0.07 & -1.35$\pm$0.03 & -1.62$\pm$0.08 & 1 \\
NGC 6705 & 8.95$\pm$0.07 & 8.28$\pm$0.12 & +0.07$\pm$0.05 & +0.07$\pm$0.07 & 7 \\
NGC 6715 & 4.86$\pm$0.03 & 4.30$\pm$0.03 & -1.25$\pm$0.06 & -1.44$\pm$0.07 & 1\\
NGC 6791 & 9.78$\pm$0.09 & 8.77$\pm$0.09 & +0.47$\pm$0.04 & +0.47$\pm$0.04 & 8\\
NGC 6819 & 8.41$\pm$0.04 & 7.64$\pm$0.04 & +0.07$\pm$0.03 & +0.09$\pm$0.03 & 9 \\
NGC 7078 & 2.78$\pm$0.01 & 2.18$\pm$0.01 & -2.12$\pm$0.04 & -2.33$\pm$0.02 &  1\\
NGC 7789 & 8.47$\pm$0.02 & 7.61$\pm$0.02 & -0.04$\pm$0.05 & -0.04$\pm$0.05 & 10\\
\enddata
\tablerefs{
(1) \citet{carretta2009}; (2) \citet{randich2003}; (3) \citet{yong2005};
(4) \citet{jacobson2009}; (5) \citet{pancino2010}; (6) \citet{friel2010}; (7) \citet{gonzalez2000}; (8) \citet{carretta2007}; (9) \citet{bragaglia2001};
(10) \citet{tautvaisiene2005}.}
\end{deluxetable}

\clearpage

\begin{deluxetable}{lcccccc}
\tablecaption{Mean, standard deviation, 90th and 10th percentile values of the metallicity distribution of each field.\label{tbl5}}
\tablewidth{0pt}
\tablehead{
\colhead{ID} & \colhead{$R (kpc)$} & \colhead{$\langle$[Fe/H]$\rangle$} & \colhead{$\sigma_{[Fe/H]}$} & \colhead{[Fe/H]$_{90th}$} & \colhead{[Fe/H]$_{10th}$} & \colhead{N$_*$}
}
\startdata
Bar & 0.3 & -0.40$\pm$0.07 & 0.35$\pm$0.07 & --0.25$\pm$0.12 & --1.25$\pm$0.30 & 373\\
LMC0512-6648 &  3.0 & --0.58$\pm$0.09 & 0.47$\pm$0.07 & --0.09$\pm$0.09 & --1.38$\pm$0.28 & 130 \\
LMC0514-6503 &  4.4 & --0.64$\pm$0.10 & 0.54$\pm$0.11 & --0.09$\pm$0.10 & --1.71$\pm$0.44 & 110 \\
LMC0513-6333 &  5.7 & --0.54$\pm$0.11 & 0.48$\pm$0.07 & --0.01$\pm$0.08 & --1.57$\pm$0.17 &  69\\
LMC0422-7100 &  5.9 & --0.43$\pm$0.09 & 0.39$\pm$0.14 & --0.09$\pm$0.07 & --1.09$\pm$0.48 &  63\\
LMC0513-6159 &  7.1 & --1.03$\pm$0.13 & 0.64$\pm$0.10 & --0.27$\pm$0.16 & --2.17$\pm$0.25 & 65\\
LMC0649-7110 &  7.1 & --0.74$\pm$0.14 & 0.62$\pm$0.15 & --0.10$\pm$0.15 & --1.79$\pm$0.14 & 24\\
LMC0512-6100 &  7.9 & --1.04$\pm$0.30 & 0.70$\pm$0.25 &--0.42$\pm$0.15 & --1.96$\pm$0.47 & 12\\
LMC0356-7100 &  8.3 & --0.87$\pm$0.23 & 0.56$\pm$0.16 & --0.48$\pm$0.08 & --2.07$\pm$0.25 & 12\\
LMC0511-6000 & 8.8 &--0.99$\pm$0.37 & 0.20$\pm$0.10 & --0.71$\pm$0.10 &--1.63$\pm$0.47 &  6 \\
LMC0714-7100 & 9.2 &--0.85$\pm$0.22 & 0.35$\pm$0.23 & --0.71$\pm$0.07 & --1.75$\pm$0.14 & 7\\
\enddata
\end{deluxetable}

\begin{deluxetable}{cccc}
\tabletypesize{\scriptsize}
\tablecaption{Velocity dispersion for each metallicity bin.\label{tbl6}}
\tablewidth{0pt}
\tablehead{
\colhead{[Fe/H]} & \colhead{Age (Gyr)} & \colhead{N$_*$} & \colhead{$\sigma_{V} (km s^{-1})$}
}
\startdata
--0.25 to +0.5 & $<$2 & 117 & 20.9$\pm$1.2\\
--0.75 to --0.25 & 2-6 & 212 & 21.8$\pm$0.9\\
--1.5 to --0.75 & 6-10 & 122 & 26.2$\pm$1.1\\
--3.5 to --1.5 & $>$10 & 45 & 26.8$\pm$1.4\\
\enddata
\end{deluxetable}

\clearpage

\begin{deluxetable}{ccc}
\tabletypesize{\scriptsize}
\tablecaption{Velocity dispersion as a function of radius.\label{tbl7}}
\tablewidth{0pt}
\tablehead{
\colhead{R (kpc)} & \colhead{N$_*$} & \colhead{$\sigma_{V} (km s^{-1})$}
}
\startdata
Bar & 374 & 25.4$\pm$0.2\tablenotemark{a}\\
2.5--3.5 & 130 & 28.6$\pm$1.5\\
3.5--5 & 172 & 23.6$\pm$0.4\\
5--7 & 138 & 21.1$\pm$1.2\\
7--10\arcdeg & 56 & 24.0$\pm$1.1\\
\enddata
\tablenotetext{a}{Value obtained from \citet{cole2005} data following the same procedure than for our sample. This value is similar to that obtained in the original work of 24.7$\pm$0.4 km s$^{-1}$.}
\end{deluxetable}

\end{document}